 \documentclass{emulateapj}

\usepackage{graphicx}
\usepackage{epsfig}
\usepackage{times}
\usepackage{natbib}
\usepackage{url}
\usepackage{color}
\usepackage{amssymb,amsmath}

\shorttitle{Electric Currents in Active Regions}
\shortauthors{T\"or\"ok et al.}

\begin{document}

\title{Distribution of electric currents in solar active regions}

\author{
T. T\"or\"ok$^{1}$,
J.~E. Leake$^{2}$,
V.~S. Titov$^{1}$,
V. Archontis$^{3}$,
Z. Miki\'c$^{1}$,  
M.~G. Linton$^{4}$ , 
K. Dalmasse$^{5}$,
G. Aulanier $^{5}$,
B. Kliem $^{6}$}
\affil{$1$ Predictive Science, Inc., 9990 Mesa Rim Rd., Ste. 170, San Diego, CA 92121, USA}
\affil{$2$ College of Science, George Mason University, 4400 University Drive, Fairfax, VA 22030, USA}
\affil{$3$ School of Mathematics and Statistics, University of St. Andrews, North Haugh, St. Andrews, Fife, KY16 9SS, UK}
\affil{$4$ U.S. Naval Research Lab, 4555 Overlook Ave., SW Washington, DC 20375, USA}
\affil{$5$ LESIA, Observatoire de Paris, CNRS, UPMC, Univ. Paris Diderot, 5 place Jules Janssen, 92190 Meudon, France}
\affil{$6$ Institut f\"ur Physik und Astronomie, Universit\"at Potsdam, Karl-Liebknecht-Str. 24-25, 14476 Potsdam, Germany}

\begin{abstract}
There has been a long-lasting debate on the question of whether or not electric currents in solar active regions are neutralized. That is, whether or not the main (or direct) coronal currents connecting the active region polarities are surrounded by shielding (or return) currents of equal total value and opposite direction. Both theory and observations are not yet fully conclusive regarding this question, and numerical simulations have, surprisingly, barely been used to address it. Here we quantify the evolution of electric currents during the formation of a bipolar active region by considering a three-dimensional magnetohydrodynamic simulation of the emergence of a sub-photospheric, current-neutralized magnetic flux rope into the solar atmosphere. We find that a strong deviation from current neutralization develops simultaneously with the onset of significant flux emergence into the corona, accompanied by the development of substantial magnetic shear along the active region's polarity inversion line. After the region has formed and flux emergence has ceased, the strong magnetic fields in the region's center are connected solely by direct currents, and the total direct current is several times larger than the total return current. These results suggest that active regions, the main sources of coronal mass ejections and flares, are born with substantial net currents, in agreement with recent observations. Furthermore, they support eruption models that employ pre-eruption magnetic fields containing such currents.
         
\end{abstract}

\keywords{magnetohydrodynamics (MHD); Sun: magnetic fields; Sun: corona; Sun: coronal mass ejections (CMEs)}

\section{Introduction}
\label{s:intro}
The energy required to power solar flares and CMEs is stored in current-carrying magnetic fields in the corona. Active regions (ARs), the main source regions of eruptions, carry a total electric current of $\sim$ 1\,TA \citep[e.g.,][]{wilkinson92}, which is commonly inferred from applying Amp\`ere's law, ${\bf j} = (\nabla \times {\bf B})/\mu_0$, to photospheric vector magnetograms. Since such data are hampered by limited resolution and various uncertainties \citep[e.g,][]{wiegelmann06}, it is not yet well understood how AR currents are distributed.

The observations indicate that the currents in magnetically well-isolated ARs are {\em balanced} to a very good approximation, i.e., the total current, $I$, calculated by integrating the vertical current density, $j_z$, over the {\em whole} photospheric AR extension vanishes, as expected from $\nabla \cdot {\bf j}=0$ \citep[e.g.,][]{georgoulis12}. What remains controversial is to which extent the currents are {\em neutralized}, meaning that $I$ calculated over a {\em single} AR polarity vanishes as well. Full neutralization requires the main (or direct) currents which connect the AR polarities to be surrounded by shielding (or return) currents of equal total strength and opposite direction \citep[see, e.g., Figure\,1 in][]{melrose95}. Both observations and theoretical considerations are not yet fully conclusive regarding the existence or amount of return currents in ARs, which has led to an ongoing debate \citep[e.g.,][]{parker96,melrose96,georgoulis12}.

\begin{figure}[t]
\centering
\includegraphics[width=1.\linewidth]{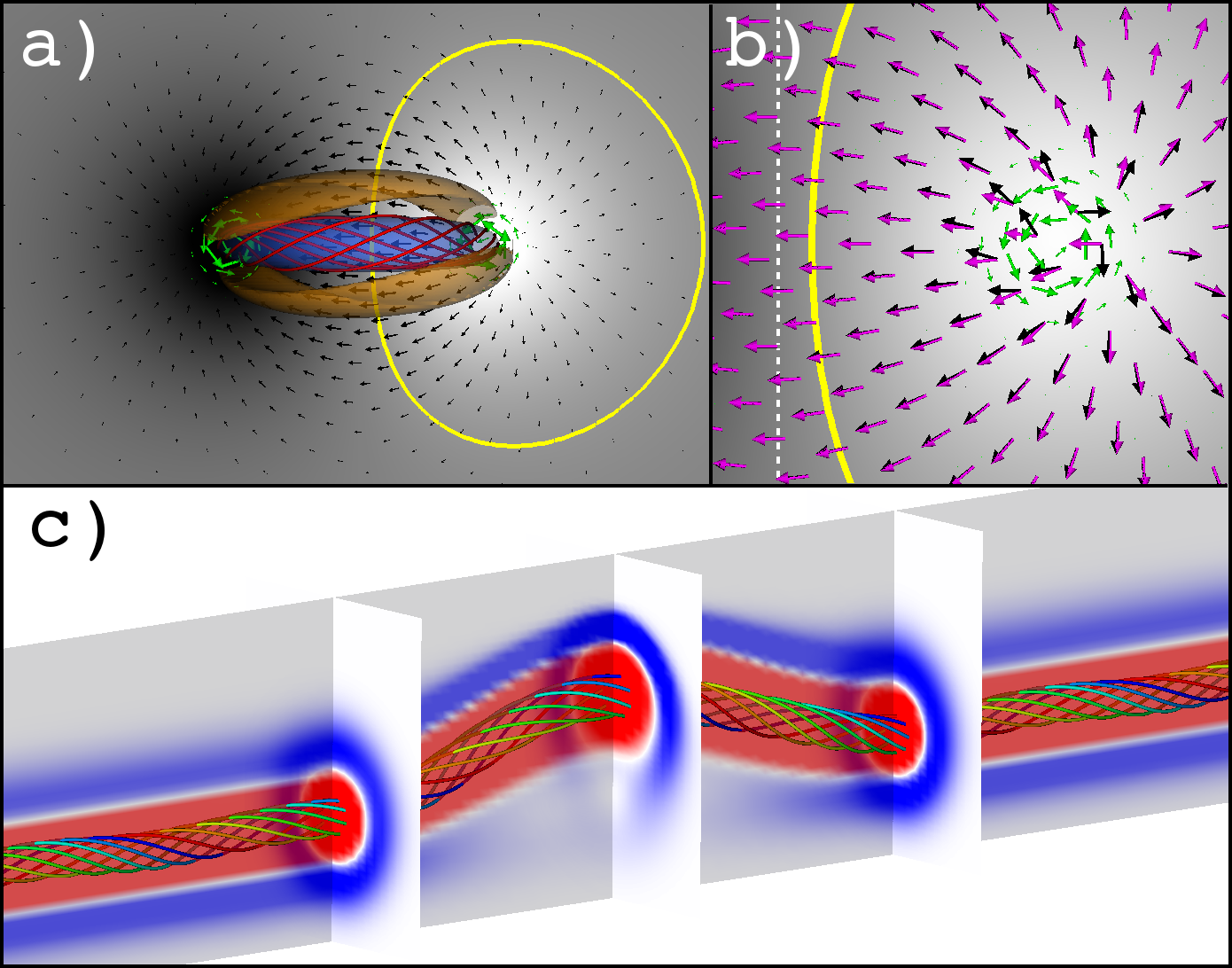}
\caption{
Electric currents in two numerical models of AR formation. (a) Bipolar AR containing a flux rope (red field lines) created by photospheric vortex flows (green arrows). Black (white) colors outline negative (positive) vertical magnetic fields. Currents are visualized by transparent iso-surfaces of $\alpha = ({\bf j} \cdot {\bf B})/{B^2}$, with $\alpha = -2$ (blue; direct current) and 0.65 (orange; return current). Black arrows show horizontal field components. The yellow line is an example path for calculating the total current in one polarity (see text). (b) Zoom into AR center, showing additionally the initial horizontal potential field (magenta arrows) and the polarity inversion line (white dotted line). (c) Buoyant flux rope from the simulation investigated in this Letter, before it emerges through the photosphere. Red (blue) colors outline direct (return) currents; field lines show the flux rope core.}
\label{fig:fg1}
\end{figure}

AR currents are believed to be formed by two main mechanisms: (i) the stressing of the coronal magnetic field by photospheric and sub-surface flows \citep[e.g.,][]{klimchuk92} and (ii) the emergence of current-carrying flux from the solar interior into the corona \citep[e.g.,][]{leka96}. At first glance, mechanism (i) is expected to produce neutralized currents. To illustrate this, we show in Fig.\,\ref{fig:fg1}(a,b) a simple AR model created from a bipolar potential field by photospheric vortex flows \citep{amari96,torok03,aulanier05,torok13}. Such an isolated, symmetric system must be current-balanced. To see if it is also neutralized, we calculate $I=\oint_C {\bf B} \cdot \mathrm{d}{\bf l}$ in {\em one} AR polarity along a photospheric path $C$ that runs fully outside and sufficiently far from the vortex flows. Since the horizontal field components along $C$ do not change much during the twisting (Figure~\ref{fig:fg1}(b)) and the initial field is current-free, $I$ remains close to zero at all times, i.e., the generated currents remain nearly neutralized. However, as shown by \cite{torok03}, net currents develop in the system if the vortices are close enough to each other to also shear the magnetic field at the polarity inversion line (PIL). The resulting handedness is the same as in the core of the flux rope, i.e., the sheared flux carries direct current. Some recent observations, based on high-resolution vector magnetograms, indeed suggest the presence of substantial net currents in ARs with strong shear along their main PIL \citep[][]{ravindra11,georgoulis12}.

As for mechanism (ii), it is believed that flux ropes rising through the convection zone are magnetically well-isolated \citep{fan09a}, which implies that the currents they carry are well-neutralized (Figure\,\ref{fig:fg1}(c)). Whether or not the neutralization breaks down when such flux ropes emerge into the corona has not yet been investigated systematically. \cite{ravindra11} analyzed a case of strong flux emergence and found it to be associated with the development of strong net currents and strong shear at the PIL. \cite{longcope00} suggested, based on a simplified analytical model, that return currents may even completely remain below the corona during the emergence of magnetically isolated flux tubes.

\begin{figure*}[t]
\centering
\includegraphics[width=1.\linewidth]{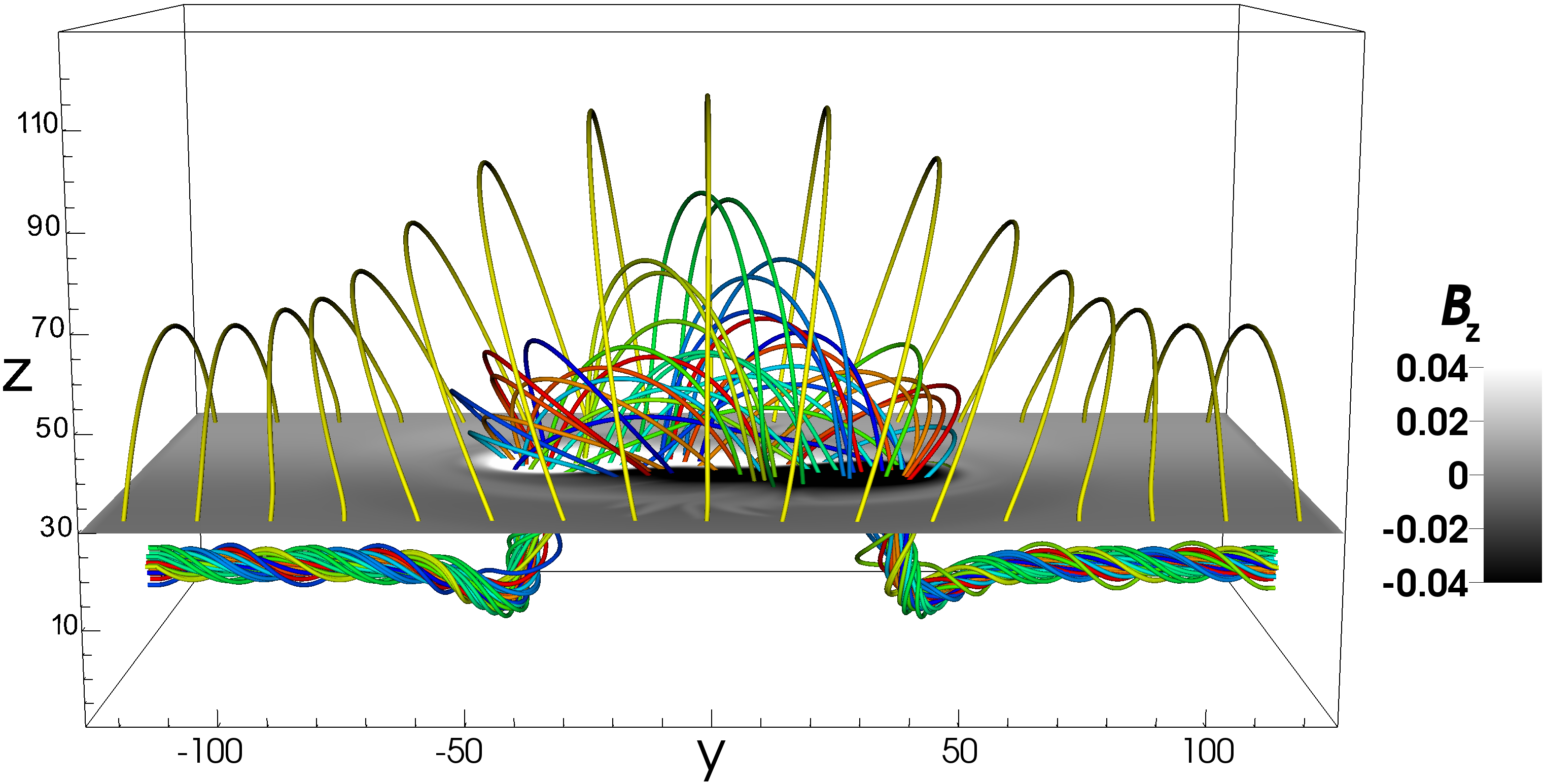}
\caption
{Perspective view on magnetic field lines of the emerging flux rope (random colors) and the ambient coronal dipole field (yellow; drawn for $z>30$) at $t=170$. The ``magnetogram'' $B_z(x,y,z=30)$ is shown in grayscale.}
\label{fig:fg2}
\end{figure*}

Improving our understanding of the current distribution in ARs is particularly important for theoretical and numerical models of solar eruptions. Many CME simulations \citep[e.g,][]{roussev03,torok05,manchester08,lugaz11,torok11a} employed the analytical coronal flux rope configuration developed by \cite{titov99}, which does not contain return currents. Other investigations used coronal field models constructed through flux rope insertion and numerical relaxation \citep{vanballegooijen04}, which are similarly dominated by direct currents, to represent the source-region field prior to an eruption \citep[e.g.,][]{bobra08,savcheva12a}. Based on the assumption that AR currents are neutralized, it has been argued, however, that such configurations are not suitable for CME modeling, as the inclusion of return currents may inhibit their eruption \citep[see a summary in][]{forbes10}.

Since theory and observations are not yet conclusive, MHD simulations can be used as a viable tool to address the question of current neutralization in ARs. Surprisingly, while the development of return currents has been reported in simulations where ARs were produced by photospheric flows \citep[e.g.,][]{aulanier05,delannee08}, the {\em amount} of current neutralization was quantified in such simulations only by \cite{torok03}. To the best of our knowledge, this has not yet been done for ARs produced in flux emergence simulations, which is the purpose of this Letter.

\section{Numerical Setup}
\label{sec:num}
The simulation analyzed here is identical to the run ``SD'' in \cite{leake13} (hereafter L13), except for a shift in the $z$ coordinate and slightly different boundary and wave damping conditions, which do not affect the system evolution noticeably. It uses the standard, Cartesian setup for the emergence of a buoyant magnetic flux rope into a stratified, plane-parallel atmosphere in hydrostatic equilibrium \citep{fan01}. In contrast to previous simulations, where often a field-free corona was considered, the flux rope here emerges into a pre-existing magnetic arcade (Figure\,\ref{fig:fg2}). We refer the reader to L13 for details, here we only note that (i) the dimensionless lengths, times, magnetic field strengths, current densities, and total currents shown below are normalized by 170\,km, 25\,s, 1200\,G, 0.56\,Am$^{-2}$, and 0.016\,TA, respectively; (ii) the height range $20<z<30$ ($0<z<10$ in L13) corresponds to the photosphere/chromosphere layer (PCL); (iii) the initial magnetic field consists of a horizontal sub-photospheric flux rope that runs along the $y$ direction and a background dipole field that is translationally invariant along the rope axis. The axis is placed at $z=8$ and the field strength at it is set to 6000\,G. The dipole field is much weaker, so initially the flux rope currents are almost perfectly neutralized. The rope is made buoyant by a localized, internal density perturbation applied around $y=0$. 

\begin{figure}[b]
\centering
\includegraphics[width=0.63\linewidth]{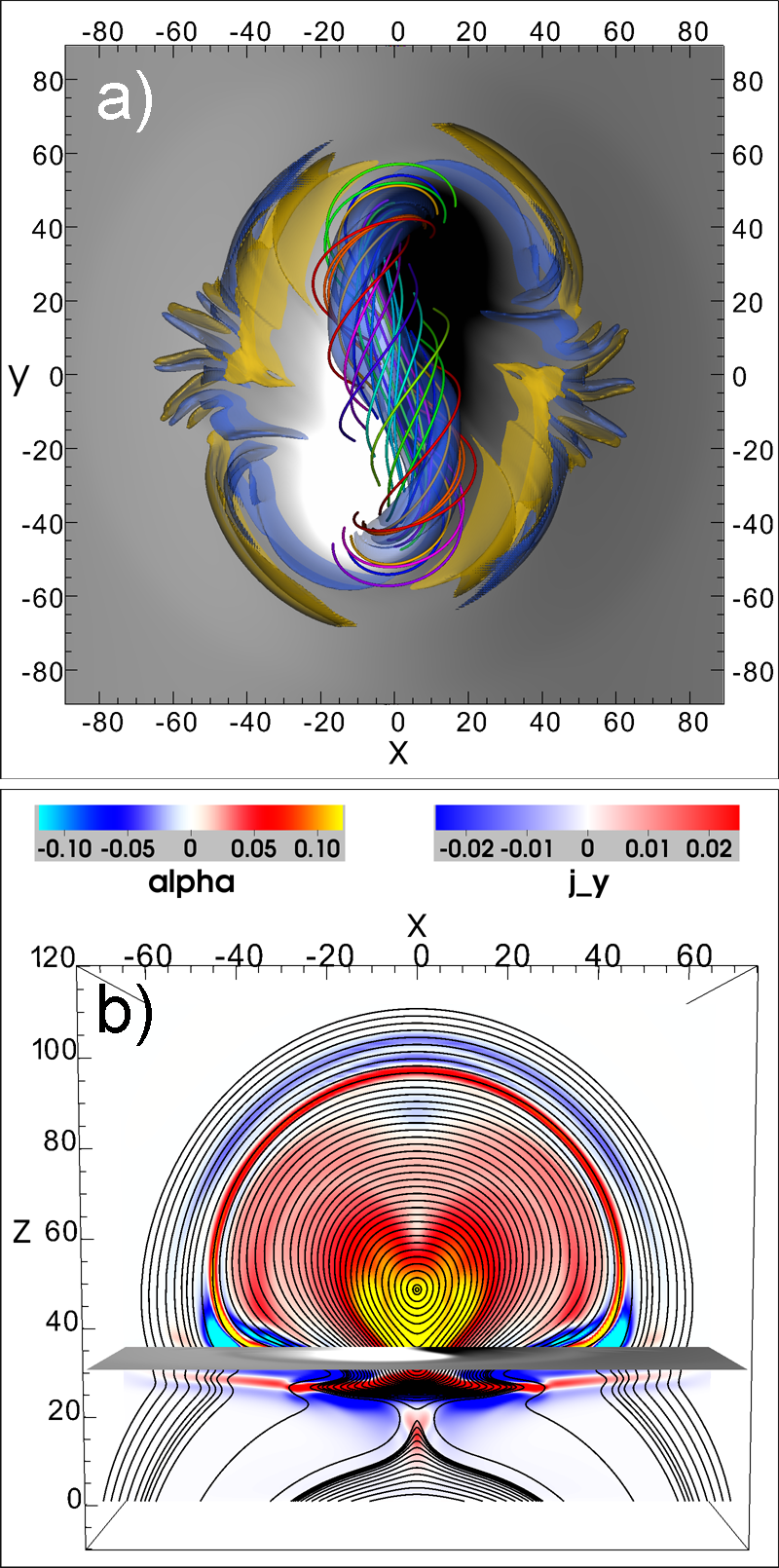}
\caption{(a) Top view of the system shown in Figure\,\ref{fig:fg2} (without arcade field lines). Direct (return) currents are visualized by a blue (orange) transparent iso-surface of $\alpha=0.12$ (-0.08). Note that the flux rope is right-handed ($\alpha>0$), while the flux rope in Figure\,\ref{fig:fg1}(a) is left-handed ($\alpha<0$). (b) Perspective view along the rope axis, showing field lines of $(B_x,0,B_z)$ and color-scales of $\alpha$ (for $z>30$) and $j_y$ (for $z<30$) in the plane $\{y=0\}$ ($j_y$ is used to visualize the current direction for $z<30$ since the field is far from a force-free state there).
}
\label{fig:fg3}
\end{figure}

\section{Results}
\label{sec:res}
The subsequent evolution is very similar to previous flux emergence simulations (see L13). Here we only show the magnetic configuration at the time when there is no more significant flux emergence into the atmosphere and a bipolar AR containing a stable flux rope has formed in the corona (Figure\,\ref{fig:fg2}). The eruption of the rope, observed in previous simulations \citep{manchester04,archontis08c}, is inhibited here by the stabilizing dipole field, the orientation of which was chosen to minimize magnetic reconnection with the emerging flux. Figure\,\ref{fig:fg3} shows that the strongest AR currents are located above the PIL and exhibit a sigmoidal shape when viewed from above. Return currents are present but are rather narrow and located at the AR edges, while the AR center and the flux rope contain only direct currents. Note that the quantity $\alpha$ represents the direct and return currents reasonably well, since the coronal configuration evolves toward a force-free state as the emergence of flux into the atmosphere slows down (see Figure\,13 in L13).

In order to quantify the AR currents, we calculate $I=\int\,j_z\,\mathrm{d}x\,\mathrm{d}y$ at the top of the PCL ($z=30$). Integration over the whole AR shows that the total current is balanced at all times, as expected. To check the amount of current neutralization, we restrict the integration to the positive AR polarity, $B_z (z=30) > 0$. The emerging flux rope has right-handed twist, so the total direct (return) current, $I_d$ ($I_r$), is obtained by integrating $j_z(z=30)>0\,(<0)$ over this polarity.
 
\begin{figure*}[t]
\centering
\includegraphics[width=1.\linewidth]{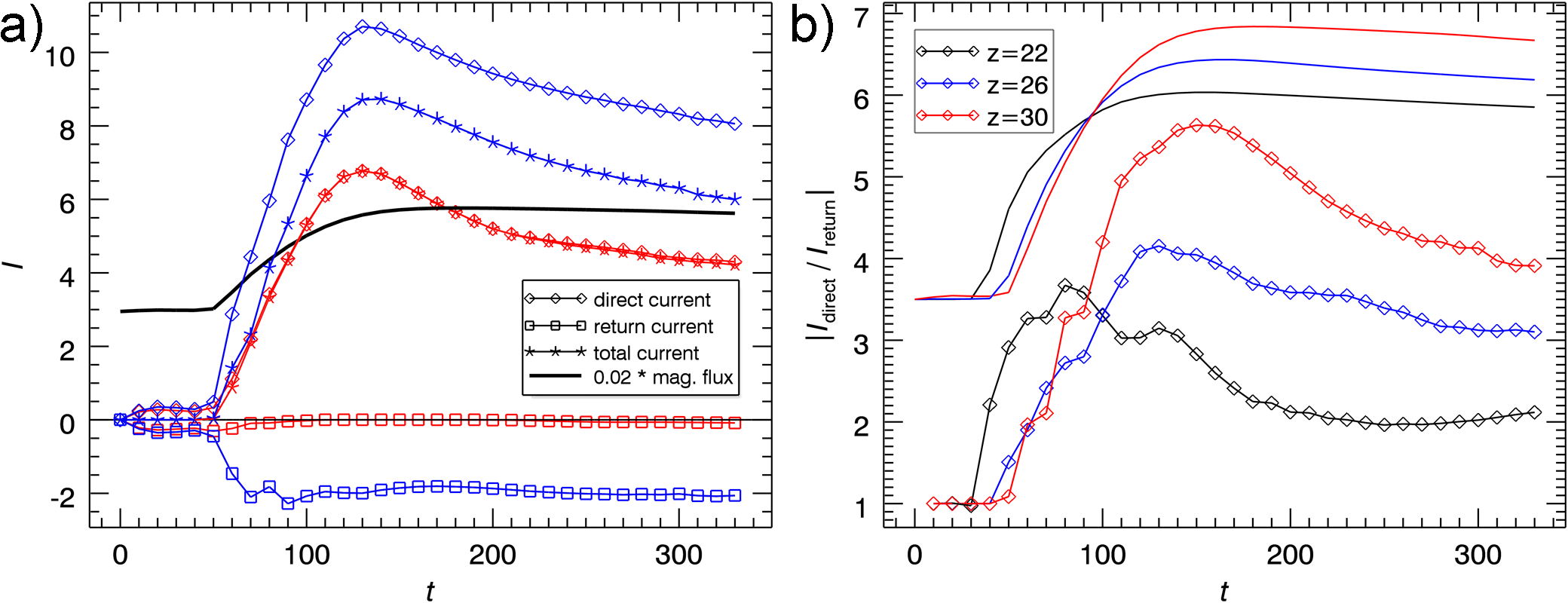}
\caption
{Electric currents integrated over the positive polarity region, $B_z(z)>0$. (a) Integration at $z=30$ over the whole polarity (blue symbols) and over the polarity center ($B_z > B_{z_{\mathrm max}}/3$; red symbols). The total positive magnetic flux is shown by a black line (scaled to fit into the plot). (b) Ratio of total direct and total return current integrated over the whole positive polarity at heights $z=22$, 26, and 30 (red, blue, and black diamonds, respectively). The total positive magnetic flux at these heights is shown by solid lines of the same color, scaled to the same initial value.}
\label{fig:fg4}
\end{figure*}

Figure\,\ref{fig:fg4}(a) shows the evolution of $I_d$, $I_r$, $I=I_d+I_r$ (blue symbols), and the total positive magnetic flux (black curve). The initial flux is non-zero due to the presence of the background dipole field. Early in the evolution ($t \lesssim 50$) there is very little flux emergence and the currents remain small and almost perfectly neutralized. No significant shear develops along the PIL during this phase. Strong emergence starts at $t \approx 50$, accompanied by a rapid increase of the currents, and ceases at $t \approx 160$. $I_r$ saturates at $t \approx 100$, while $I_d$ increases until $t \approx 130$ and slowly decreases afterwards. 

Figure\,\ref{fig:fg4}(b) shows the ratio $|I_d/I_r|$ (blue diamonds). The total direct current starts to exceed the total return current from the onset of strong emergence and remains several times larger during the whole evolution. The same pattern can be found if $I_d$ and $I_r$ are computed deeper in the PCL (at $z=22$ and 26), with somewhat smaller values of $|I_d/I_r|$.

The red symbols in Figure\,\ref{fig:fg4}(a) show $I_d$, $I_r$, and $I$ in the center of the positive polarity, where the strongest magnetic fields are located. The integration area was defined by the ad-hoc condition $B_z(z=30,t)>B_{z_{\mathrm max}}(z=30,t)/3$ (see the black contour lines in Figure\,\ref{fig:fg5}(a,c)). It can be seen that $I_r$ in the AR center drops to zero shortly after the onset of strong emergence, i.e, the strongest AR fields become connected solely by direct currents as the emergence proceeds. This is visualized in Figure\,\ref{fig:fg5}(a--d). At $t=55$, right after the onset of strong emergence, the direct and return currents are still quite compact and more or less equally distributed within each polarity of the forming AR. No significant shear along the PIL has yet developed. The PILs of $B_z$ and $j_z$ are very different, indicating that the system is far from a force-free state at this height and time. As the emergence proceeds the picture changes considerably. At $t=170$, when flux emergence has ceased and strong shear along the PIL has developed, two \textsf{J}-shaped regions of strong direct current occupy the AR center, while the much weaker and narrower return currents are located solely in the AR's periphery. The PILs of $B_z$ and $j_z$ in the AR center now coincide, indicating that the coronal configuration has evolved to an approximately force-free state. This pattern persists during the remaining evolution of the system, except that as the AR polarity centers separate, the current concentrations between them progressively narrow, plausibly causing the decrease of $I_d$ after $t \approx 130$.

\section{Discussion}
\label{sec:dis}
We quantified the amount of electric current neutralization in bipolar solar ARs by considering an MHD simulation of the emergence of a current-neutralized magnetic flux rope from the solar interior into the corona. We find that a rapid and strong deviation from current-neutralization occurs simultaneously with the onset of significant flux emergence. The emergence process is accompanied by the development of strong magnetic shear along the AR's PIL. By the end of the emergence phase $|I_d|$ is several times larger than $|I_r|$ for the model parameters considered here, in reasonable agreement with the ratios obtained from observed data by \cite{ravindra11} and \cite{georgoulis12}. The strong magnetic fields in the AR center are connected solely by direct currents, while the weaker and narrower return currents reside in the AR's periphery. In order to assess the role of shielding on the strength of the return currents, we repeated the simulation using a three times stronger dipole field, and also compared it with emergence into a field-free corona (run ``ND'' in L13). The deviation from current neutralization is strong in all cases and increases with the ambient field strength, opposite to expectation if shielding were dominant. These results suggests that: 

\vspace{0.18cm}
1. ARs are born with substantial net currents, in agreement with recent observations \citep{ravindra11,georgoulis12}.

\vspace{0.18cm}
2. Coronal flux rope models that neglect return currents \citep[e.g.,][]{titov99, su11} are a valid representation of pre-eruption configurations on the Sun. Indeed, simulations that use such models reproduce important eruption characteristics (e.g., rise profiles and morphological evolution) in very good quantitative agreement with the observations \citep[e.g.,][]{torok05,williams05,schrijver08a,kliem10,kliem12,kliem13}.

\vspace{0.18cm}

The question arises of how fully neutralized sub-photospheric currents transform into strongly non-neutralized coronal currents during flux emergence. This transformation is not trivial, since (i) the current paths become highly complex during the rise and emergence of the flux rope, (ii) only a fraction of the sub-photospheric currents enter the corona, and (iii) new currents may develop as a result of the shearing and converging flows associated with emerging flux ropes \citep{manchester04,archontis08b} or of the transport of twist from below the surface via torsional Alfv\'en waves (\citealt{longcope00}; \citealt{fan09}; L13). 

\begin{figure*}[t]
\centering
\includegraphics[width=1.\linewidth]{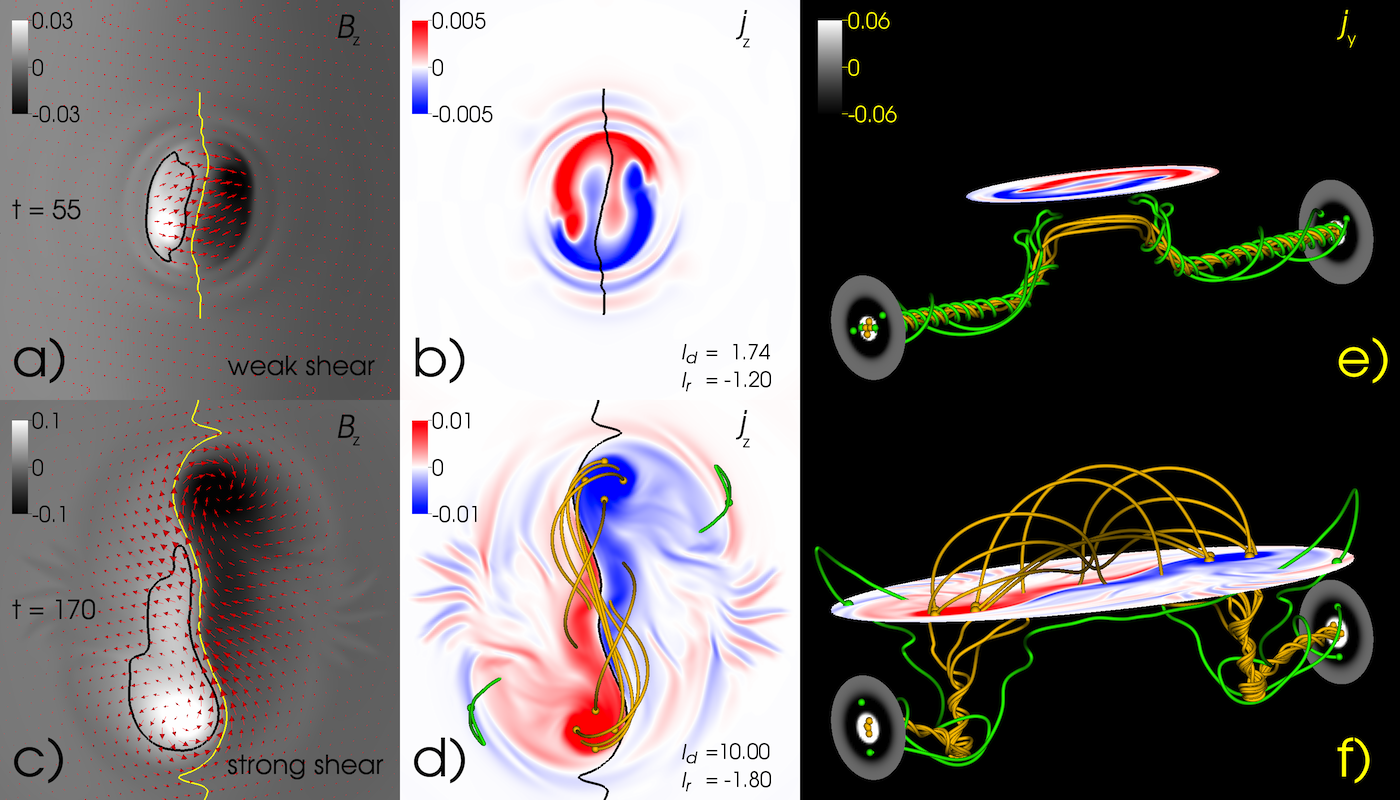}
\caption
{Development of net currents in the corona. The top shows an early state of emergence, when little shear has developed and the currents are relatively well neutralized. The bottom shows a mature state of emergence, after strong shear has developed and the currents are strongly non-neutralized. (a--d) $B_z$ and $j_z$ at $z=30$. On the left, red arrows, yellow and black lines outline $(B_x,B_y,0)$, $B_z=0$, and contours of $B_{z_{\mathrm max}}/3$, respectively. On the right, $B_z=0$ is drawn in black. The current field lines in (d) are the same as in (f). (e--f) Oblique view showing $j_z(z=30)$, field lines of {\bf j}, and vertical slices of $j_y$ at $y= \pm 75$, with white (black) regions outlining the locations of sub-photospheric direct (return) currents. Current field lines start at $y= \pm 75$ in (e) and at $z=30$ in (f), and are terminated at $y= \pm 75$ in both panels. Orange (green) lines start in regions of direct (return) current. Green current field lines in (e) have short-circuited, i.e., they connect to the direct current region in the same flux rope leg.}
\label{fig:fg5}
\end{figure*}

The complexity of the problem calls for a detailed investigation beyond the scope of this Letter. A preliminary analysis indicates that during the flux pile-up that occurs when the rising flux rope approaches the photosphere, return currents located at the top of the rope are pushed aside by subjacent direct currents. Moreover, some of them short-circuit with adjacent direct currents (Figure\,\ref{fig:fg5}(e)), which supports this process. It appears that most of the return currents thus relocated to the periphery of the emergence area never enter the corona (otherwise the flux emergence would start with an increase of the return current). Figure\,\ref{fig:fg5}(f) shows that the direct currents that occupy the AR center after emergence are rooted in the center of the sub-photospheric flux rope. This suggests that they emerge bodily, rather than being produced by shearing flows.

It also needs to be studied how parameters such as the initial flux rope twist and diameter (relative to the PCL width), and the structure and strength of the pre-existing coronal field affect the final current distribution.

The results presented here refer to newly emerging flux, in particular to emerging ARs. While the most powerful eruptions tend to arise from relatively young and compact ARs, many filament eruptions and CMEs originate within or between decaying ARs characterized by dispersed photospheric flux distributions \citep{martin73,tang87}. The corresponding pre-eruption configurations (typically filament channels) are believed to be formed and energized primarily by persistent shear flows and flux cancellation at PILs \citep[e.g.,][]{vanballegooijen89,martens01,green09,green11}, rather than by newly emerging flux. Their magnetic structure has been modeled using two complementary approaches.

The flux rope insertion method \citep{vanballegooijen04} yields static models by inserting a flux rope into the potential-field extrapolation of an observed magnetogram and subsequently relaxing the coronal field numerically \citep[e.g.,][]{bobra08, savcheva09, su11}. After relaxation, the currents in these models tend to be strongly non-neutralized \citep[see Figure\,3 in][]{kliem13}. In particular, the use of a potential ambient field neglects the possible introduction of return currents by localized shearing flows. Nevertheless, the models have been very successful in representing stable as well as unstable fields of dispersed and decaying ARs. 

MHD simulations involving flux cancellation have produced fully dynamic models of such ARs \citep[e.g.,][]{linker03, amari03b, titov08, aulanier10}. These simulations impose various combinations of photospheric shearing, converging flows, and field diffusion on an initial potential field; thus, possible return currents resulting from the shearing are kept. Figure\,12 in \cite{titov08} and Figure\,7 in \cite{aulanier10} show that return currents are produced in these models, but they appear to be too weak to neutralize the strong direct currents that form above the PIL. For example, the simulation in \cite{titov08} yields $|I_d/I_r| \approx 3$, rather similar to the values shown in Figure\,\ref{fig:fg4}. 

All three models (flux emergence, flux rope insertion, and flux cancellation) appear to produce strongly non-neutralized current configurations, though their quantitative differences have yet to be investigated systematically. The same is true for configurations produced by localized vortex flows (Figure\,\ref{fig:fg1}(a)) if these flows extend close to the PIL. Moreover, these configurations are morphologically similar, typically consisting of a low-lying, sigmoidal layer of strong and concentrated currents and a flux rope with weaker and more diffuse current above it \citep{torok03,archontis09,aulanier10,savcheva12a}. This suggests that young eruptive ARs and mature CME source regions have analogous current distributions, dominated by net currents located close to the PIL, though this conjecture needs to be substantiated by further analysis of numerical models and observations. 

Finally, all models show an association between the presence of net currents and magnetic shear along the PIL, regardless of whether the currents are produced by horizontal photospheric flows or emerge bodily into the corona. While the shear at the PIL is causal for the current to be non-neutralized in the vortex-driven case, it may not be causal in the emergence process. The exact nature and validity range of the relationship requires further study as well.

\acknowledgments
We thank the referee for helpful comments and Brian Welsch for stimulating discussions. The contributions of T.T., V.S.T., and Z.M. were supported by NASA's HTP, LWS, and SR\&T programs. J.E.L and M.G.L. were supported by NASA/LWS. M.G.L. received support also from the ONR 6.1 program. The simulation was performed under grant of computer time from the D.o.D. HPC Program. B.K. was supported by the DFG. V.A. acknowledges support through the IEF-272549 grant.


\begin{thebibliography}{50}
\expandafter\ifx\csname natexlab\endcsname\relax\def\natexlab#1{#1}\fi

\bibitem[{{Amari} {et~al.}(2003){Amari}, {Luciani}, {Aly}, {Mikic}, \&
  {Linker}}]{amari03b}
{Amari}, T., {Luciani}, J.~F., {Aly}, J.~J., {Mikic}, Z., \& {Linker}, J. 2003,
  \apj, 595, 1231

\bibitem[{{Amari} {et~al.}(1996){Amari}, {Luciani}, {Aly}, \&
  {Tagger}}]{amari96}
{Amari}, T., {Luciani}, J.~F., {Aly}, J.~J., \& {Tagger}, M. 1996, \apjl, 466,
  L39

\bibitem[{{Archontis}(2008)}]{archontis08b}
{Archontis}, V. 2008, Journal of Geophysical Research (Space Physics), 113, 3

\bibitem[{{Archontis} \& {Hood}(2009)}]{archontis09}
{Archontis}, V., \& {Hood}, A.~W. 2009, \aap, 508, 1469

\bibitem[{{Archontis} \& {T{\"o}r{\"o}k}(2008)}]{archontis08c}
{Archontis}, V., \& {T{\"o}r{\"o}k}, T. 2008, \aap, 492, L35

\bibitem[{{Aulanier} {et~al.}(2005){Aulanier}, {D{\'e}moulin}, \&
  {Grappin}}]{aulanier05}
{Aulanier}, G., {D{\'e}moulin}, P., \& {Grappin}, R. 2005, \aap, 430, 1067

\bibitem[{{Aulanier} {et~al.}(2010){Aulanier}, {T{\"o}r{\"o}k}, {D{\'e}moulin},
  \& {DeLuca}}]{aulanier10}
{Aulanier}, G., {T{\"o}r{\"o}k}, T., {D{\'e}moulin}, P., \& {DeLuca}, E.~E.
  2010, \apj, 708, 314

\bibitem[{{Bobra} {et~al.}(2008){Bobra}, {van Ballegooijen}, \&
  {DeLuca}}]{bobra08}
{Bobra}, M.~G., {van Ballegooijen}, A.~A., \& {DeLuca}, E.~E. 2008, \apj, 672,
  1209

\bibitem[{{Delann{\'e}e} {et~al.}(2008){Delann{\'e}e}, {T{\"o}r{\"o}k},
  {Aulanier}, \& {Hochedez}}]{delannee08}
{Delann{\'e}e}, C., {T{\"o}r{\"o}k}, T., {Aulanier}, G., \& {Hochedez}, J.-F.
  2008, \solphys, 247, 123

\bibitem[{{Fan}(2001)}]{fan01}
{Fan}, Y. 2001, \apjl, 554, L111

\bibitem[{{Fan}(2009{\natexlab{a}})}]{fan09a}
---. 2009{\natexlab{a}}, Living Reviews in Solar Physics, 6, 4

\bibitem[{{Fan}(2009{\natexlab{b}})}]{fan09}
---. 2009{\natexlab{b}}, \apj, 697, 1529

\bibitem[{{Forbes}(2010)}]{forbes10}
{Forbes}, T. 2010, in Heliophysics: Space Storms and Radiation: Causes and
  Effects, ed. C.~J. {Schrijver} \& G.~L. {Siscoe} (Cambridge (UK): Cambridge
  University Press), 159

\bibitem[{{Georgoulis} {et~al.}(2012){Georgoulis}, {Titov}, \&
  {Miki{\'c}}}]{georgoulis12}
{Georgoulis}, M.~K., {Titov}, V.~S., \& {Miki{\'c}}, Z. 2012, \apj, 761, 61

\bibitem[{{Green} \& {Kliem}(2009)}]{green09}
{Green}, L.~M., \& {Kliem}, B. 2009, \apjl, 700, L83

\bibitem[{{Green} {et~al.}(2011){Green}, {Kliem}, \& {Wallace}}]{green11}
{Green}, L.~M., {Kliem}, B., \& {Wallace}, A.~J. 2011, \aap, 526, A2

\bibitem[{{Kliem} {et~al.}(2010){Kliem}, {Linton}, {T{\"o}r{\"o}k}, \&
  {Karlick{\'y}}}]{kliem10}
{Kliem}, B., {Linton}, M.~G., {T{\"o}r{\"o}k}, T., \& {Karlick{\'y}}, M. 2010,
  \solphys, 266, 91

\bibitem[{{Kliem} {et~al.}(2013){Kliem}, {Su}, {van Ballegooijen}, \&
  {DeLuca}}]{kliem13}
{Kliem}, B., {Su}, Y., {van Ballegooijen}, A., \& {DeLuca}, E. 2013, \apj, 779, 129

\bibitem[{{Kliem} {et~al.}(2012){Kliem}, {T{\"o}r{\"o}k}, \&
  {Thompson}}]{kliem12}
{Kliem}, B., {T{\"o}r{\"o}k}, T., \& {Thompson}, W.~T. 2012, \solphys, 281, 137

\bibitem[{{Klimchuk} \& {Sturrock}(1992)}]{klimchuk92}
{Klimchuk}, J.~A., \& {Sturrock}, P.~A. 1992, \apj, 385, 344

\bibitem[{{Leake} {et~al.}(2013){Leake}, {Linton}, \&
  {T{\"o}r{\"o}k}}]{leake13}
{Leake}, J.~E., {Linton}, M.~G., \& {T{\"o}r{\"o}k}, T. 2013, \apj, 788, 99

\bibitem[{{Leka} {et~al.}(1996){Leka}, {Canfield}, {McClymont}, \& {van
  Driel-Gesztelyi}}]{leka96}
{Leka}, K.~D., {Canfield}, R.~C., {McClymont}, A.~N., \& {van Driel-Gesztelyi},
  L. 1996, \apj, 462, 547

\bibitem[{{Linker} {et~al.}(2003){Linker}, {Miki\'c}, {Lionello}, {Riley},
  {Amari}, \& {Odstrcil}}]{linker03}
{Linker}, J.~A., {Miki\'c}, Z., {Lionello}, R., et al.
2003, Phys. of Plasmas, 10, 1971

\bibitem[{{Longcope} \& {Welsch}(2000)}]{longcope00}
{Longcope}, D.~W., \& {Welsch}, B.~T. 2000, \apj, 545, 1089

\bibitem[{{Lugaz} {et~al.}(2011){Lugaz}, {Downs}, {Shibata}, {Roussev}, {Asai},
  \& {Gombosi}}]{lugaz11}
{Lugaz}, N., {Downs}, C., {Shibata}, K., et al.
2011, \apj, 738, 127

\bibitem[{{Manchester} {et~al.}(2004){Manchester}, {Gombosi}, {DeZeeuw}, \&
  {Fan}}]{manchester04}
{Manchester}, IV, W., {Gombosi}, T., {DeZeeuw}, D., \& {Fan}, Y. 2004, \apj,
  610, 588

\bibitem[{{Manchester} {et~al.}(2008){Manchester}, {Vourlidas}, {T{\'o}th},
  {Lugaz}, {Roussev}, {Sokolov}, {Gombosi}, {De Zeeuw}, \&
  {Opher}}]{manchester08}
{Manchester}, IV, W.~B., {Vourlidas}, A., {T{\'o}th}, G., et al.
2008, \apj, 684, 1448

\bibitem[{{Martens} \& {Zwaan}(2001)}]{martens01}
{Martens}, P.~C., \& {Zwaan}, C. 2001, \apj, 558, 872

\bibitem[{{Martin}(1973)}]{martin73}
{Martin}, S.~F. 1973, \solphys, 31, 3

\bibitem[{{Melrose}(1995)}]{melrose95}
{Melrose}, D.~B. 1995, \apj, 451, 391

\bibitem[{{Melrose}(1996)}]{melrose96}
---. 1996, \apj, 471, 497

\bibitem[{{Parker}(1996)}]{parker96}
{Parker}, E.~N. 1996, \apj, 471, 489

\bibitem[{{Ravindra} {et~al.}(2011){Ravindra}, {Venkatakrishnan}, {Tiwari}, \&
  {Bhattacharyya}}]{ravindra11}
{Ravindra}, B., {Venkatakrishnan}, P., {Tiwari}, S.~K., \& {Bhattacharyya}, R.
  2011, \apj, 740, 19

\bibitem[{{Roussev} {et~al.}(2003){Roussev}, {Forbes}, {Gombosi}, {Sokolov},
  {DeZeeuw}, \& {Birn}}]{roussev03}
{Roussev}, I.~I., {Forbes}, T.~G., {Gombosi}, T.~I., et al.
2003, \apjl, 588, L45

\bibitem[{{Savcheva} \& {van Ballegooijen}(2009)}]{savcheva09}
{Savcheva}, A., \& {van Ballegooijen}, A. 2009, \apj, 703, 1766

\bibitem[{{Savcheva} {et~al.}(2012){Savcheva}, {van Ballegooijen}, \&
  {DeLuca}}]{savcheva12a}
{Savcheva}, A.~S., {van Ballegooijen}, A.~A., \& {DeLuca}, E.~E. 2012, \apj,
  744, 78

\bibitem[{{Schrijver} {et~al.}(2008){Schrijver}, {Elmore}, {Kliem},
  {T{\"o}r{\"o}k}, \& {Title}}]{schrijver08a}
{Schrijver}, C.~J., {Elmore}, C., {Kliem}, B., {T{\"o}r{\"o}k}, T., \& {Title},
  A.~M. 2008, \apj, 674, 586

\bibitem[{{Su} {et~al.}(2011){Su}, {Surges}, {van Ballegooijen}, {DeLuca}, \&
  {Golub}}]{su11}
{Su}, Y., {Surges}, V., {van Ballegooijen}, A., {DeLuca}, E., \& {Golub}, L.
  2011, \apj, 734, 53

\bibitem[{{Tang}(1987)}]{tang87}
{Tang}, F. 1987, \solphys, 107, 233

\bibitem[{{Titov} \& {D{\'e}moulin}(1999)}]{titov99}
{Titov}, V.~S., \& {D{\'e}moulin}, P. 1999, \aap, 351, 707

\bibitem[{{Titov} {et~al.}(2008){Titov}, {Mikic}, {Linker}, \&
  {Lionello}}]{titov08}
{Titov}, V.~S., {Mikic}, Z., {Linker}, J.~A., \& {Lionello}, R. 2008, \apj,
  675, 1614

\bibitem[{{T{\"o}r{\"o}k} \& {Kliem}(2003)}]{torok03}
{T{\"o}r{\"o}k}, T., \& {Kliem}, B. 2003, \aap, 406, 1043

\bibitem[{{T{\"o}r{\"o}k} \& {Kliem}(2005)}]{torok05}
---. 2005, \apjl, 630, L97

\bibitem[{{T{\"o}r{\"o}k} {et~al.}(2011){T{\"o}r{\"o}k}, {Panasenco}, {Titov},
  {Miki{\'c}}, {Reeves}, {Velli}, {Linker}, \& {De Toma}}]{torok11a}
{T{\"o}r{\"o}k}, T., {Panasenco}, O., {Titov}, V.~S., et al.
2011, \apjl, 739, L63

\bibitem[{{T{\"o}r{\"o}k} {et~al.}(2013){T{\"o}r{\"o}k}, {Temmer}, {Valori},
  {Veronig}, {van Driel-Gesztelyi}, \& {Vr{\v s}nak}}]{torok13}
{T{\"o}r{\"o}k}, T., {Temmer}, M., {Valori}, G., et al.
2013, \solphys, 286, 453

\bibitem[{{van Ballegooijen}(2004)}]{vanballegooijen04}
{van Ballegooijen}, A.~A. 2004, \apj, 612, 519

\bibitem[{{van Ballegooijen} \& {Martens}(1989)}]{vanballegooijen89}
{van Ballegooijen}, A.~A., \& {Martens}, P.~C.~H. 1989, \apj, 343, 971

\bibitem[{{Wiegelmann} {et~al.}(2006){Wiegelmann}, {Inhester}, \&
  {Sakurai}}]{wiegelmann06}
{Wiegelmann}, T., {Inhester}, B., \& {Sakurai}, T. 2006, \solphys, 233, 215

\bibitem[{{Wilkinson} {et~al.}(1992){Wilkinson}, {Emslie}, \&
  {Gary}}]{wilkinson92}
{Wilkinson}, L.~K., {Emslie}, A.~G., \& {Gary}, G.~A. 1992, \apjl, 392, L39

\bibitem[{{Williams} {et~al.}(2005){Williams}, {T{\"o}r{\"o}k}, {D{\'e}moulin},
  {van Driel-Gesztelyi}, \& {Kliem}}]{williams05}
{Williams}, D.~R., {T{\"o}r{\"o}k}, T., {D{\'e}moulin}, P., {van
  Driel-Gesztelyi}, L., \& {Kliem}, B. 2005, \apjl, 628, L163

\end{thebibliography}
\end{document}